\documentclass[a4paper,10pt]{article}
\usepackage{amsthm}
\newtheorem{lemma}{Lemma}

\newtheorem{cor}{Corollary}
\newtheorem{theorem}{Theorem}
\usepackage{graphicx}
\usepackage{graphicx,amssymb}
\usepackage{graphicx,amssymb,amsmath}
\usepackage{hyperref}
\usepackage{epsfig, color, graphicx}
\usepackage{enumerate}
\usepackage[margin=1.0in]{geometry}
\usepackage{amssymb}

\newtheorem{nc'}{Necessary Condition}

\title{On colouring point visibility graphs}
\author{Ajit Arvind Diwan \thanks{Department of Computer Science and Engineering,
        Indian Institute of Technology Bombay, {\tt aad@cse.iitb.ac.in}}
        \and
        Bodhayan Roy \thanks{Faculty of Informatics, Masaryk University, Brno, Czech Republic,
         {\tt b.roy@fi.muni.cz}}
         \footnote{
        Most of the work was done when the author was affiliated to the 
        Department of Computer Science and Engineering, Indian Institute of Technology Bombay.}
        }

\begin{document}
\thispagestyle{empty}
\maketitle
 
\begin{abstract}
In this paper we show that it can be decided in polynomial time whether or not the visibility graph of a given
point set is $4$-colourable, and such a $4$-colouring, if it exists, can also be constructed in polynomial time.
We show that the problem of 
deciding whether the visibility graph of a point set is $5$-colourable, 
is NP-complete.
We give an example of a point visibility graph that has chromatic number $6$ while its clique number is only $4$.
\end{abstract}
\section{Introduction} 
The visibility graph is a fundamental structure studied in the field of computational geometry  
and geometric graph theory \cite{bcko-cgaa-08,g-vap-07}. 
 Some of the early applications of visibility graphs included computing 
Euclidean shortest paths in the presence of obstacles \cite{lw-apcf-79} and decomposing 
two-dimensional shapes into clusters \cite{sh-ddsg-79}. 
Here, we consider problems concerning the colouring of visibility graphs.
$\\ \\$
Let $P$ be a set of points $\{p_1, p_2, \ldots, p_n \}$ in the plane. Two points $p_i$ and $p_j$ of $P$ are said to be
\emph{mutually visible} if there is no third point $p_k$ on the line segment joining $p_i$ and $p_j$. Otherwise, $p_i$ and $p_j$ 
are said to be mutually \emph{invisible}.
The \emph {point visibility graph} (denoted as PVG) $G(V,E)$ of $P$ is defined as follows. 
The set $V$ of vertices contains a vertex $v_i$ for every
 point $p_i$ in $P$. The set $E$ contains an undirected edge $v_iv_j$ if and only if the corresponding points $p_i$ and $p_j$
are mutually visible \cite{prob-ghosh}.
Point visibility graphs have been studied in the contexts of 
construction \cite{cgl-pgd-85,Edelsbrunner:1986:CAL}, 
recognition \cite{pvg-card,prob-ghosh,pvg-tcs,pvg-np-hard}, partitioning \cite{pvg_part},
connectivity \cite{viscon-wood-2012},
chromatic number and clique number \cite{penta-2013,kpw-ocnv-2005,p-vgps-2008}. 
$\\ \\$
A graph is said to be \emph{$k$-colourable} if each vertex of the graph can be assigned a colour, so that no 
two adjacent vertices are assigned the same colour, and the total number of distinct colours assigned to the vertices
is at most $k$.
K\'{a}ra et al characterized PVGs that are $2$-colourable and $3$-colourable \cite{kpw-ocnv-2005}. 
 It was not known whether the 
  chromatic number of a PVG can be found in polynomial time.
  In Section \ref{sec5col} we show that the
  problem of deciding whether a PVG is $k$-colourable, for $k \geq 5$, is NP-complete.
$\\ \\$
K\'{a}ra et al also asked whether there is a function $f$ such that for
every point visibility graph $G$, $\chi (G) \leq f(\omega (G))$ \cite{kpw-ocnv-2005}? \label{prob3}
They presented a family of 
PVGs that have their chromatic number lower bounded by an exponential function of their 
clique number.  
Their question was answered by Pfender, showing that for a PVG with $\omega (G) = 6$, $\chi (G)$ can be arbitrarily large \cite{p-vgps-2008}.
However, it is not known whether the chromatic number of a PVG is bounded, if its clique number is only $4$ or $5$.
In another related paper, Cibulka et al showed that PVGs of point sets $S$ such that there is no convex pentagon with vertices in $S$ and no other point of $S$ lying 
in the pentagon, 
might have arbitrarily large clique numbers \cite{penta-2013}.
In this direction, K\'{a}ra et al showed that there is a PVG $G$ with $\omega (G) = 4$ and $\chi (G) = 5$ \cite{kpw-ocnv-2005}. 
In Section \ref{secexample2} we construct a PVG $G'$ with $\omega (G') = 4$ and $\chi (G') = 6$.
\section{Four-colouring} \label{sec4col} 
In this section, we provide a polynomial-time algorithm to decide if the PVG of a given point set is 4-colourable, and construct a 4-colouring if it exists.
Consider a finite set $P$ of $n$ points in the Euclidean plane.
We start with a brief overview of our algorithm:
\begin{enumerate} [(i)]
 \item Check if $P$ is 3-colourable. If $P$ is 3-colourable then construct the 3-colouring and terminate. 
 \item Find a convex hull vertex of $P$ that forms a $K_4$ with three other vertices. Delete this convex hull vertex from $P$.
 Repeat this step until there is no such convex hull vertex.
 \item There are at most eight possible colourings of the reduced $P$. Check if any of these colourings is valid. If none of 
 then are valid then output ``NO'' and terminate.
 \item Consider each of the valid colourings and progressively add the deleted points to $P$, in the reversed order of their deletion.
 With each addition, colour the added point and check if the colouring is valid. If it is not valid then output ``NO'' and terminate.
\end{enumerate}

In the next section we provide a proof of correctness and analysis of our algorithm.
\subsection{Correctness of the algorithm}
The algorithm begins with checking for 3-colourability. This can be done in polynomial time due to K\'{a}ra et al \cite{kpw-ocnv-2005}.
We present the following lemma and theorem from  K\'{a}ra et al verbatim without proof. Note that in Lemma \ref{kara1}, $V(P)$ being planar 
actually means that $V(P)$ drawn on $P$ is plane. 
\begin{figure} 
\begin{center}
\centerline{\hbox{\psfig{figure=66.pdf,width=0.40 \hsize}}}
\caption{(a) }
\label{figocta}
\end{center}
\end{figure}
 \lemma \label{kara1} 
\textbf{(K\'{a}ra et al \cite{kpw-ocnv-2005})}
 Let $P$ be a point set. Then $V(P )$ is planar if and only if at least one of
the following conditions hold:
\begin{enumerate} [(a)]
\item all the points in $P$ are collinear,
\item all the points in $P$ , except for one, are collinear,
\item  all the points in $P$ are collinear, except for two non-visible points,
\item all the points in $P$ are collinear, except for two points $v$, $w \in P$ , such that the 
line-segment $vw$ does not intersect the line-segment that contains $P \setminus \{ v, w \}$,
\item $V(P)$ is the drawing of the octahedron shown in Figure \ref{figocta}.
\end{enumerate}
 
\theorem 
\label{karatheo}
\textbf{(K\'{a}ra et al \cite{kpw-ocnv-2005})}
  Let $P$ be a finite point set. Then the following are equivalent: 
  \begin{enumerate} [(i)]
\item  $\chi (V(P )) \leq 3$,
\item  $P$ satisfies conditions (a), (b), (c) or (e) in Lemma \ref{kara1},
\item  $V(P )$ has no $K_4$ subgraph.
\end{enumerate}

\emph{ If the algorithm finds $P$ to be 3-colourable, then it produces a 3-colouring and terminates.
Suppose that the algorithm finds that $P$ is not 3-colourable.
Then the algorithm proceeds to the next step.
It deletes any convex hull vertex that sees three mutually visible points in the rest of $P$.
It continues this process till no such convex hull vertex is left. We call the resultant point set the \emph{reduced set} $P_r$. 
The set $P_r$ can be obtained from $P$ in $O(n^4)$ time. 
We have the following lemma.}
$\\ \\$
 \lemma \label{lemrestr}
 $P$ is 4-colourable only if $P_r$ is 4-colourable, and given a 4-colouring of $P_r$, it can be found in polynomial time 
 if it is a 4-colouring of $P$ restricted to $P_r$
 
\proof
The contrapositive of the first part is easy to see, since the PVG of $P_r$ is an induced subgraph of the PVG of $P$.
$\\ \\$
Consider the deleted points of $P$ in the reverse order of their deletion. 
Since each deleted point sees a $K_3$ in the remaining points of $P$,
its colour must be uniquely determined by the remaining points of $P$. Given a 4-colouring of $P_r$, 
we can add the deleted points in the reverse
order of their deletion, colour them and check if the colouring is valid. 
It takes $O(n^3)$ time for each point to locate its corresponding $K_3$,
and $O(n^2)$ time to check if the colouring is valid.
So, the total procedure takes $O(n^4)$ time.
\qed
$\\ \\$
Now the algorithm checks if $P_r$ is 4-colourable.
First it checks if $P_r$ is 3-colourable. According to the characterization in Theorem \ref{karatheo}, this can be achieved in polynomial time.
Furthermore, we have the following lemmas.
 \lemma \label{k3}
 A reduced set must contain a $K_3$.
 
\proof
A reduced set is obtained only after progressively deleting all convex hull vertices which see a $K_3$. 
So, after the final deletion step, the $K_3$ which the deleted point saw must remain.
\qed
 \lemma \label{unq3col}
 If a reduced set is 3-colourable, then it requires three colours, and has a unique 3-colouring.
 
\proof
By Theorem \ref{karatheo}, any 3-colourable PVG must be of the forms (a), (b), (c) or (e) of Lemma \ref{kara1}.
Among these, the PVGs of the forms (b), (c) or (e), require 3-colours and have unique 3-colourings. 
A reduced set can never be of the form (a), i.e. all collinear points, because by Lemma \ref{k3} every reduced set must contain a $K_3$.
\qed
\begin{figure} 
\begin{center}
\centerline{\hbox{\psfig{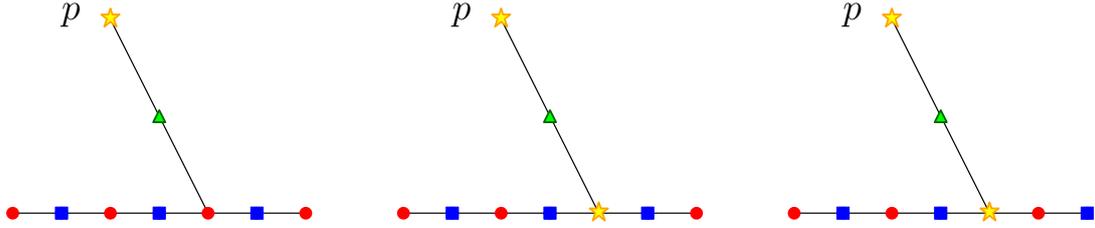}}}
\caption{Three 4-colourings of point $p$ added to reduced set of type (b) in Lemma  \ref{kara1}. 
}
\label{figcol1} 
\end{center}
\end{figure}
\begin{figure} 
\begin{center}
\centerline{\hbox{\psfig{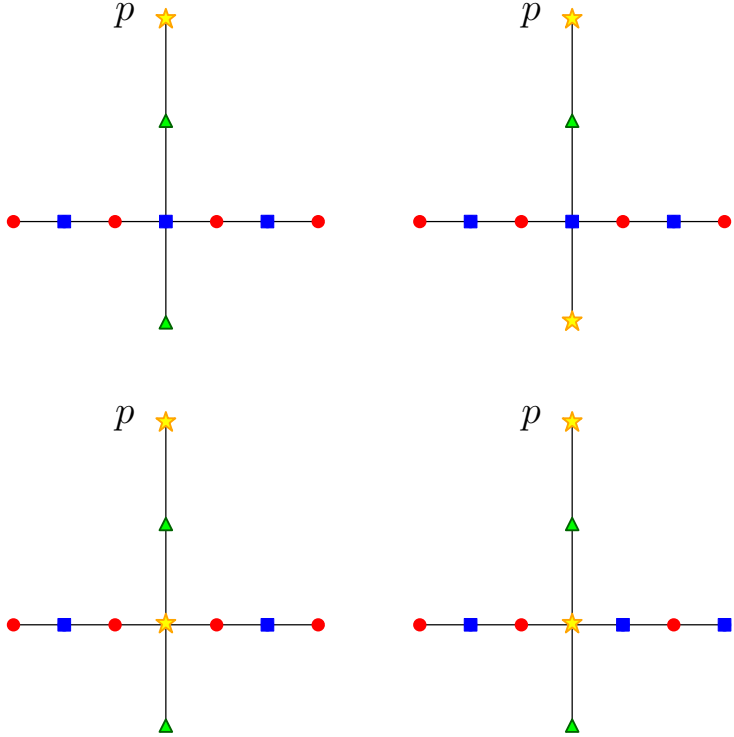}}}
\caption{Four colourings of point $p$ added to reduced set of type (c) in Lemma  \ref{kara1}. 
}
\label{figcol2} 
\end{center}
\end{figure}
 \lemma \label{cont4col}
 A reduced 3-colourable set is no more 3-colourable if its last deleted point is added to it. It then requires four colours and can have no more than a constant number of 
 4-colourings. 
 
\proof
Suppose that the last deleted point $p$ is added back to the reduced set.
Due to Lemma \ref{unq3col} the reduced set can only be of three types.
If the reduced set is of type (b) in Lemma  \ref{kara1}, then the three colourings in Figure \ref{figcol1} are the only possibilities.
If the reduced set is of type (c) in Lemma  \ref{kara1}, then the four colourings in Figure \ref{figcol2} are the only possibilities.
If the reduced set is of type (e) in Lemma  \ref{kara1}, then if has only a constant number of points and hence a constant number of 4-colourings.
\qed
$\\ \\$
If the algorithm finds $P_r$ to be 3-colourable, then it adds the last deleted point to $P_r$, 
and due to Lemma \ref{cont4col} constructs a constant number of 4-colourings in polynomial time. For each 4-colouring of $P_r$, the algorithm then 
reintroduces the deleted points of $P$ progressively in the reversed order of their deletion, and by Lemma \ref{lemrestr}, constructs 
a 4-colouring of $P$ if it exists, in $O(n^3)$ time. Suppose that the algorithm finds that $P_r$ is not 3-colourable. Then it checks
whether $P_r$ is 4-colourable. Now, we describe the structure of reduced sets that are 4-colourable but not 3-colourable.
$\\ \\$
Let $p_x$ be a vertex of the convex hull of $P_r$.
All the other points of $P_r$  lie on rays emanating from $p_x$.
Here we consider only \emph{open rays} emanating from $p_x$, i.e. rays that do not contain their initial point $p_x$. 
Let $r_1, r_2, \ldots r_k$ be the rays emanating from $p_x$ in the clockwise order, 
with the rays $r_1$ and $r_k$ respectively
being tangents from $p_x$ to the convex hull of $P \setminus \{ p_x \}$.
Let $q_i$ be the closest point on $r_i$ to $p_x$.
We call the path $(q_1, q_2, \ldots, q_k)$
the \emph{frontier} of $p_x$. 
The points $q_1, q_2, \ldots, q_k$ are called the \emph{frontier points} of $p_x$.
The points $q_2, q_3, \ldots, q_{k-1}$ are called the \emph{internal frontier points} of $p_x$.
A continuous subpath $(q_i, q_{i+1}, q_{i+2})$ of the frontier is said to be a \emph{convex triple} 
(or, \emph{concave}) when $q_{i+2}$,
 lies to the right (respectively, left) of the ray $\overrightarrow{q_iq_{i+1}}$.
If the continuous subpath is a straight line-segment then it is said to be a \emph{straight triple}.
The frontier might have concave, straight and convex triples. 
We have the following lemmas.
\begin{figure} 
\begin{center}
\centerline{\hbox{\psfig{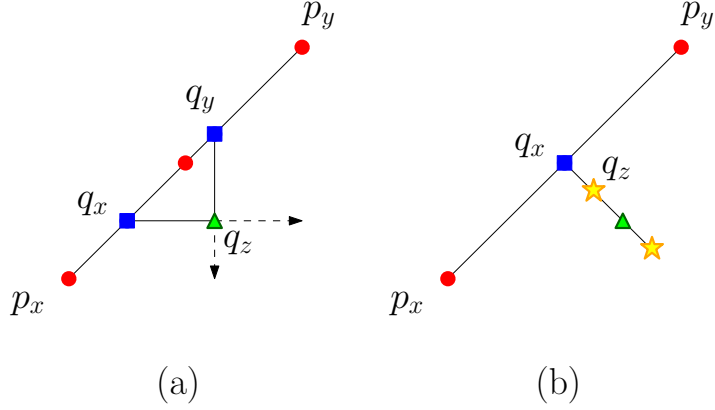}}}
\caption{
(a) The convex hull points $p_x$ and $p_y$ have more than one point between them. 
(b) The convex hull points $p_x$ and $p_y$ have exactly one point $q_x$ between them. 
}
\label{fignopoint}
\end{center}
\end{figure}
 \lemma \label{noconv}
 The following holds for each convex hull vertex of $P_r$:
 \begin{enumerate} [(a)]
  \item There is no concave triple in its frontier.
  \item If there is a convex triple in its frontier then the ray containing the convex vertex has at least two points.
 \end{enumerate}
 
\proof 
If three consecutive points of its frontier are concave with respect to $p_x$, then they together form a $K_4$.
If three consecutive points of its frontier are convex, then they do not form a $K_3$ if and only if there is a blocker
on the ray containing the convex vertex that prevents the first points of its two neighbouring rays from seeing each other.
\qed

 \lemma \label{lemintfront}
 If a reduced set is not 3-colourable then each of its convex hull vertices has an internal frontier point.
 
\proof
Suppose that a convex hull vertex $p_x$ of the reduced set $P_r$ does not have an internal frontier point.
Then the rest of the points of $P_r$ must be lying on only two rays emanating from $p_x$.
Denote these two rays as $r_1$ and $r_2$ and the points of $P_r$ on them farthest from $p_x$ as $p_1$ and $p_2$ respectively.
If both $r_1$ and $r_2$ each have two or more points excluding $p_x$, then $p_1$, $p_2$ and the two points preceding them form a $K_4$.
This is a contradiction, since $P_r$ is reduced and both $p_1$ and $p_2$ are convex hull vertices of $P_r$.
If at least one of $r_1$ and $r_2$ has only one point excluding $p_x$ then $P_r$ is of the form (b) in Lemma \ref{kara1} and hence
3-colourable, a contradiction.
\qed
 \lemma \label{lemtype}
If a reduced set is not 3-colourable, then
all of its convex hull points are vertices.

\proof 
Assume on the contrary that the point set is not 3-colourable and not all the convex hull points are convex hull vertices. 
This means that there 
are two consecutive convex hull vertices $p_x$ and $p_y$ such that there is at least one 
convex hull point in the interior of the line segment joining them. 
Wlog assume that $p_x$ precedes $p_y$ in the clockwise order of points 
on the convex hull. There can be either more than one or exactly one point in the interior of $\overline{p_xp_y}$.
$\\ \\$
First consider the case where there is more than one point in the interior of $\overline{p_xp_y}$ (Figure \ref{fignopoint}(a)). Let $q_x$ 
and $q_y$ be the points among them closest to $p_x$ and $p_y$ respectively.
By Lemma \ref{lemintfront}, $p_x$ has an internal frontier point. 
Let $q_z$ be the first internal frontier point of $p_x$ following $q_x$.
Suppose that $q_z$ is not an internal frontier point of $p_y$. Then there must be a point (say, $q_u$) on $\overline{p_yq_z}$
that is an internal frontier point of $p_y$. But then there must also be an internal frontier point $q_w$ of $p_x$ on $\overline{p_xq_u}$,
contradicting the assumption that $q_z$ is the first internal frontier point of $p_x$.
So, $q_z$ must be an internal frontier point of $p_y$.
By Lemma \ref{noconv}, the frontier of $p_x$ cannot have concave triples, so that the rest of the frontier
points of $p_x$ must lie on or to the left of $\overline{q_xq_z}$.
Similarly, the frontier of $p_y$ cannot have concave triples, so that the rest of the frontier
points of $p_y$ must lie on or to the right of $\overline{q_yq_z}$.
But if there is a frontier point $q_t$ of $p_x$ on or to the left of $\overline{q_xq_z}$,
then $p_y$ must have a frontier point on $\overline{p_yq_t}$ that is to the right of $\overline{q_yq_z}$, a contradiction.
$\\ \\$
Now consider the case where there is exactly one point in the interior of $\overline{p_xp_y}$ 
and denote it as $q_x$ (Figure \ref{fignopoint}(b)). As before, we consider $q_z$ which is a frontier point of both $p_x$ and $p_y$.
As before, by Lemma \ref{noconv}, the frontier of $p_x$ cannot have concave triples, so that the rest of the frontier
points of $p_x$ must lie on or to the left of $\overline{q_xq_z}$.
Similarly, the frontier of $p_y$ cannot have concave triples, so that the rest of the frontier
points of $p_y$ must lie on or to the right of $\overline{q_xq_z}$.
This means that all of the points of $P_r$ other than $p_x$ and $p_y$ must lie on $\overline{q_xq_z}$.
But then $P_r$ is of the form (c) in Lemma \ref{kara1} and hence
3-colourable, a contradiction.
\qed
\cor \label{cor1}
If a reduced set is not 3-colourable, and its convex hull has at least four vertices, then the interior of its convex hull is not empty.
 
\proof
Suppose on the contrary the interior of the convex hull of such a reduced set $P_r$ is empty.  By Lemma \ref{lemtype} $P_r$ has no convex hull points
that are not convex hull vertices. So, all the convex hull vertices of $P_r$ see each other. Since $P_r$ has at least four
convex hull points, all of them are a part of a $K_4$. Hence, $P_r$ is not reduced, a contradiction.
\qed
\begin{figure} 
\begin{center}
\centerline{\hbox{\psfig{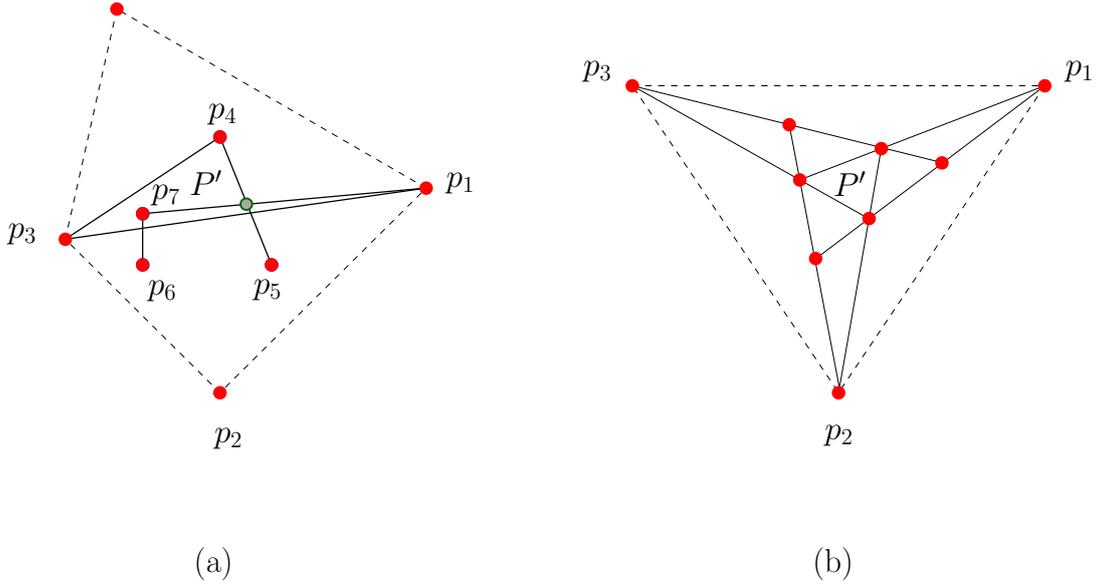}}}
\caption{(a) The intersection of $\overline {p_1p_7}$  and $\overline {p_4p_5}$ forces another point 
on $\overline {p_4p_5}$. (b) A non 3-colourable reduced set with only three convex hull vertices}
\label{figstruct}
\end{center}
\end{figure}
 \lemma \label{lemstruct}
 If a reduced set is not 3-colourable, then
  its convex hull has only three vertices.

\proof Suppose on the contrary that a reduced set $P_r$ is not 3-colourable and its convex hull has at least four vertices.
Consider the lowest vertex of the convex hull of $P_r$, and its two adjacent vertices on the convex hull of $P_r$. 
Call these three points $p_1$, $p_2$ and $p_3$ in the clockwise order (Figure \ref{figstruct}(a)). 
Denote as $P'$ the set of points other than the convex hull vertices of $P_r$. 
$\\ \\$
Since by our assumption the convex hull of 
$P_r$ has at least four vertices, by Corollary \ref{cor1}, $P'$ must be nonempty.
Suppose that all the points of $P'$ lie on $\overline {p_1p_3}$.
If there are at least two convex hull vertices above $\overline {p_1p_3}$,
then they can see two mutually visible points of $P'$, and hence $P_r$ is not reduced, a contradiction.
If there is only one convex hull vertex above $\overline {p_1p_3}$, then either
it sees $p_2$ and two mutually visible points of $P'$, which means $P_r$ is not reduced, 
or it is blocked from $p_2$, thereby making $P_r$ 3-colourable, a contradiction.
So, not all the points of $P'$ lie on $\overline {p_1p_3}$.
$\\ \\$
If all the points of $P'$ lie above $\overline {p_1p_3}$, then the lowermost point of $P'$ forms a $K_4$
with $p_1$, $p_2$ and $p_3$, a contradiction. 
Similarly, not all the points of $P'$ can lie below $\overline {p_1p_3}$ as well.
So, the convex hull of $P'$ must intersect $\overline {p_1p_3}$.
Suppose that the convex hull of $P'$ intersects $\overline {p_1p_3}$ at only one point, say $p_i$.
If there are points of $P'$ to the left or right of the ray $\overrightarrow{p_2p_i}$,
then among such points let $p_j$ be a point closest to $\overline {p_1p_3}$.  
Wlog if $p_j$ lies to the right of $\overrightarrow{p_2p_i}$ then it sees both $p_i$ and $p_1$.
But $p_2$ sees $p_1$, $p_i$ and $p_j$, so $P_r$ is not reduced, a contradiction.
Otherwise, all the points of $P'$ lie on the ray $\overrightarrow{p_2p_i}$. 
If a fourth convex hull vertex of $P_r$ above $\overline {p_1p_3}$ 
 lies on $\overrightarrow{p_2p_i}$,
then $P_r$ is 3-colourable, a contradiction. 
Otherwise, wlog let this fourth convex hull vertex
lie to the left of $\overrightarrow{p_2p_i}$.
Let $p_j$ be the point of $P'$ immediately after $p_i$
on $\overrightarrow{p_2p_i}$. Then $p_3$ forms a $K_4$
with $p_i$, $p_j$ and the fourth convex hull vertex, so $P_r$ is not reduced, a contradiction.
So, the convex hull of $P'$ must intersect $\overline {p_1p_3}$ at two points.
$\\ \\$
Let $\overline {p_4p_5}$ and $\overline {p_6p_7}$ be the segments of the convex hull of $P'$ intersecting $\overline {p_1p_3}$,
where $p_4$ and $p_5$ (respectively, $p_6$ and $p_7$) are consecutive points on the convex hull of $P'$. Also
assume that $p_4$, $p_5$, $p_6$ and $p_7$ are in the clockwise order on the convex hull of $P'$,
with none of the segments $\overline {p_1p_4}$ $\overline {p_1p_5}$
$\overline {p_3p_6}$ and $\overline {p_3p_7}$ intersecting the convex hull of $P'$.  
$\\ \\$
Consider the segment $\overline {p_3p_4}$. Suppose that $\overline {p_3p_4}$ and $\overline {p_6p_7}$ intersect.
To prevent a concave frontier of $p_3$ from forming, there must be a point of $P'$
on the intersection of $\overline {p_3p_4}$ and $\overline {p_6p_7}$. 
But that is not possible because $p_6$ and $p_7$ are consecutive points on the convex hull of $P'$.
Thus, $\overline {p_6p_7}$ must lie to the right 
of $\overrightarrow {p_3p_4}$. But then, $\overline {p_4p_5}$ and $\overline {p_1p_7}$ must intersect. So,
there must be another point of $P'$ on $\overline {p_4p_5}$, which is a contradiction to $p_4$ and $p_5$
being consecutive points on the convex hull of $P'$ (Figure \ref{figstruct}(a)).
Hence, the convex hull of $P$ can have at most three points (Figure \ref{figstruct}(b)).
\qed
$\\ \\$
Let $P_r$ be a reduced set and $p_x$ be a convex hull vertex of $P_r$.
We will henceforth refer to the four colours as red, blue, yellow and green. 
If a ray emanating from a convex hull vertex of a reduced set has only one point, we call it a \emph{small ray}.
Otherwise, 
we call it a \emph{big ray}. On a ray, the point closest to $p_x$ is called its \emph{first point}, next closest point 
to $p_x$ is called its \emph{second point} and so on.
We have the following lemmas.

 \lemma \label{noblock}
 The first point of any ray emanating from a convex hull vertex of a reduced set can block only first points of other rays from each other.
 
\proof
Consider three rays $r_i$, $r_j$ and $r_k$ emanating from a convex hull vertex $p_x$ lying in clockwise order around it. 
Let $p_2$ be the first point of $r_j$. Suppose that $p_2$ blocks $p_1$ and $p_3$ lying on $r_i$ and $r_k$ respectively,
and wlog $p_1$ is not the first point of $r_i$. Let $p_i$ be the first point of $r_i$. 
If both of the triangles $\bigtriangleup p_ip_2p_x$ and $\bigtriangleup p_2p_3p_x$ are empty, then $p_x$ forms a $K_4$ 
with $p_i$, $p_2$ and $p_r$, hence $P_r$ is not reduced.
So, at least one of the two triangles must be nonempty.
Wlog suppose that $\bigtriangleup p_1p_2p_x$ is nonempty. Let $p_4$ be a point contained in $\bigtriangleup p_ip_2p_x$
such that no other point contained in $\bigtriangleup$ is closer
to $\overline {p_1p_2}$. Then if $\bigtriangleup p_2p_3p_x$ is empty then $p_4$
forms a $K_4$ with $p_x$, $p_2$ and $p_3$. If $\bigtriangleup p_2p_3p_x$ too is nonempty then let analogously
$p_5$ be a point contained in $\bigtriangleup p_2p_3p_x$
that is a point closest to $\overline {p_1p_2}$. 
Again, $p_3$, $p_4$, $p_5$ and $p_i$ form a $K_4$, so $P_r$ is not reduced, a contradiction.
\qed

\begin{figure} 
\begin{center}
\centerline{\hbox{\psfig{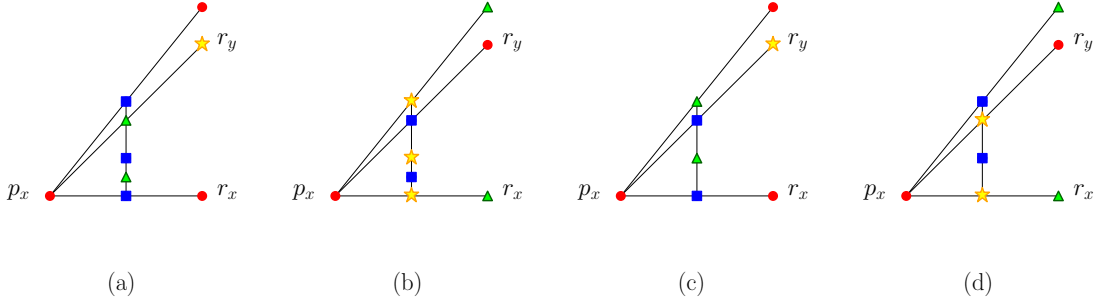}}}
\caption{
(a) The break is even and $r_x$ has a red point. 
(b) The break is even and $r_x$ does not have a red point.
(c) The break is odd and $r_x$ has a red point.
(d) The break is odd and $r_x$ does not have a red point.}
\label{fig1col}
\end{center}
\end{figure}

\lemma \label{2must}
A reduced set that is not 3-colourable must have at 
least two big rays emanating from each of its convex hull vertices.

\proof
Consider a reduced set $P_r$ that is not 3-colourable.
 Suppose that there is a convex hull vertex $p_x$ of $P_r$ such 
that only one big ray $r_x$ emanates from it. 
First suppose that $r_x$ has only two points.
By Lemma \ref{noconv}, the frontier of $p_x$ is either convex or a straight line.
If the frontier of $p_x$ is a straight line, then $P_r$ is 3-colourable, a contradiction.
So, the frontier of $p_x$ must be convex.
Suppose the first point of some small ray is a convex point in the frontier of $p_x$.
Then, the last and second-last points of the frontier of $p_x$ in the same side of $r_x$,
form a $K_4$ with the two points of $r_x$.
Among the four points forming the $K_4$, the last point of the frontier or the second point of $r_x$
is a convex hull vertex of $P_r$. Then $P_r$ is not reduced, a contradiction.
So, the first point of $r_x$ must be the only convex point in the frontier of $p_x$
Denote the two neighours of the first point of $r_x$ on the frontier of $p_a$ as $p_1$ and $p_b$.
If $p_a$ and $p_b$ see each other then they form a $K_4$ with $p_x$ and the first point of $r_x$.
Then since $p_x$ sees a $K_3$, $P_r$ is not reduced, a contradiction. 
Hence, $p_a$ and $p_b$ must be blocked from each other by the second point of $r_x$.
The frontier of $p_x$ must have more points, for otherwise $P_r$ is 3-colourable.
But now, each point of the frontier sees both the points of $r_x$.
One of the end points of the frontier must be a convex hull vertex. This end point
forms a $K_4$
with both points of $r_x$ and the second-last point of the frontier in the same side.
Hence $P_r$ is not reduced, a contradiction.
$\\ \\$
Now suppose that $r_x$ has at least three points.
If the frontier of $p_x$ has two or more points in the same side of $r_x$,
then the last and second-last points of the frontier in that side form 
a $K_4$ with the last and second-last points of $r_x$.
Since the last point of the frontier or the last point of $r_x$
is a convex hull vertex, $P_r$ is not a reduced set, a contradiction.
Then the frontier of $p_x$ can have at most one point on each side of $r_x$.
If there is no point on one side of $r_x$, then $P_r$ is 3-colourable.
So the frontier of $p_x$ must have exactly one point on each side of $r_x$.
If these two end points of the frontier do not see each other then again $P_r$ is 3-colourable.
So, the two end points of the frontier see each other. But they also see 
the last and second-last points of $r_x$, which means that $P_r$ is not reduced, a contradiction.

\qed
 \lemma \label{lem3colray}
In any 4-colouring of a reduced set that is not 3-colourable,
any big ray emanating from a convex hull vertex has exactly two colours.
\proof
Let $p_x$ be a convex hull vertex and $\{r_1, r_2, \ldots, r_k \}$ be the open rays emanating from it in the clockwise angular order.
Suppose that a big ray $r_x$ emanating from $p_x$ has three points that are assigned three different colours.
By Lemma \ref{2must}, there is another big ray with respect to $p_x$. 
Consider a big ray $r_y$ such that there is no other big ray between $r_x$ and $r_y$, and $y<x$, without loss of generality.
If $r_x$ and $r_y$ are neighbouring rays, i.e. $y=x-1$, then both of the first and second points of $r_y$ must be assigned 
the fourth colour, which is not possible.
Suppose that there is at least one small ray between $r_x$ and $r_y$.
Then the first point of the ray $r_{y+1}$ forms a $K_5$ with the first two points of $r_y$
and the second and third points of $r_x$, a contradiction.
\qed
$\\ \\$
We call the occurrence of a small ray after a big ray, or vice versa, a \emph{break}.
If there are a consecutive odd number of small rays before or after the break, it is called an \emph{odd break}.
If there are a consecutive even number of small rays before or after the break, it is called an \emph{even break}.
We have the following lemmas.

 \lemma \label{fixcol}
 A reduced set that is not 3-colourable can have at most a constant number of 4-colourings. 
\proof
By Lemma \ref{2must}, in such sets, at least two big rays emanate from every convex hull vertex.
We consider a reduced set $P_r$.
Suppose that $P_r$ is 4-colourable.
Consider a convex hull vertex $p_x$ of $P_r$.
By our assumption, at least two big rays must emanate from $p_x$.
Wlog assign red to $p_x$. Consider any big ray $r_x$ and suppose that it contains the colour red.
Due to Lemma \ref{lem3colray}, each ray can be assigned at most two colours. Wlog let the other colour of $r_x$
be blue. Observe that since $p_x$ is red, the first point of $r_x$ must be blue, the second point red and so on.
If a neighbouring ray of $r_x$ is also big, then it must have yellow and green alternatingly assigned to its points,
but either of yellow and green can be assigned to its first point. In general, till a break occurs, every alternate
big ray must contain red points. Furthermore, the second point of every such big ray must be red and red points should 
occur alternatingly on it. Hence, our initial choice of $r_x$ as a ray containing red points fixes the assignment of red 
till a break occurs.
$\\ \\$
Suppose that we have one or more consecutive big rays in which the assignment of red is fixed, and a break occurs after a big ray $r_x$.
Suppose that this break is even.
Call the first big ray occuring after the break $r_y$. Suppose that $r_x$ contains red points. Wlog let the other colour of $r_x$ be blue.
(Figure \ref{fig1col} (a)). Then the second point of $r_y$ must 
be assigned either yellow or green. Suppose that it is yellow. 
Then the colour of the only point in the first small ray occuring in the break is determined by the 
first and second points of $r_x$ and the second point of $r_y$, and it must be green.
Similarly, the next small ray gets blue. The first point of $r_y$ gets green.
Thus, yellow and green alternate throughout $r_y$ and if  
a neighbouring ray of $r_y$ is big, then it must contain a red point.
$\\ \\$
Suppose that $r_x$ does not contain any red point (Figure \ref{fig1col} (b)), and wlog the first and second points of $r_x$ are yellow and green respectively.
Then the only points of the small rays between $r_x$ and $r_y$ must be assigned blue and yellow alternately.
The first and second points of $r_y$ must be assigned blue and red respectively.
So, the assigment of red to points in rays containing an even break depends only 
on whether or not $r_x$ contains a red point. 
$\\ \\$
Suppose that a break is odd. Suppose that the break starts after $r_x$, and ends at $r_y$, both being big rays. 
Supose that $r_x$ contains a red point (Figure \ref{fig1col} (c)).
Wlog let the other colour of $r_x$ be blue.
Then the first and second points of $r_x$ must be blue and red respectively.
Then the second point of $r_y$ can be either yellow or green. Wlog let it be yellow. Then green and blue must be alternately 
assigned to the only points of the small rays in between, and the first point of $r_y$ must be blue. 
Thus, $r_y$ does not contain a red point.
$\\ \\$
Now suppose that $r_x$ does not contain any red point (Figure \ref{fig1col} (d)). 
Wlog suppose that the first and second points of $r_x$ are yellow and green respectively.
Then the second point of $r_y$ must be red, and the first points of all rays till $r_y$ must be 
assigned blue and yellow alternately.
$\\ \\$
In all cases, the points that are assigned red are fixed. Thus, the assignment of red has only two possibilities, which depend
on our initial choice of whether or not $r_x$ contains a red point. Now, by Lemma \ref{lemtype}, 
all the convex hull points of a reduced
set are also its convex hull vertices, and by Lemma \ref{lemstruct} 
it can have at most three convex hull vertices. In a 4-colouring, these three convex hull
vertices must be assigned three distinct colours. For each of these three colours, there are at most two possible assignments to the rest of the points of $P_r$.
Each assignment of three colours also fixes the assignment of the fourth colour.
This means that there are at most eight possible four colourings.
\qed
$\\ \\$
Thus, our algorithm checks if any of the eight colourings are valid 4-colourings, and then adds the deleted points back to $P_r$, assigning them their
unique colours.
Finally, we sum up our algorithm in the following theorem.
\theorem
It can be determined in polynomial time if a point set has a 4-colouring. Such a 4-colouring, if it exists, can be constructed in polynomial time.
 
\proof
If the given point set $P$ is 3-colourable then it is be identified and 3-coloured in $O(n^2)$ time due to Theorem \ref{karatheo}.
If $P$ is not 3-colourable then it is reduced to $P_r$ in $O(n^4)$ time by the method of Lemma \ref{lemrestr}.
$\\ \\$
If the reduced set $P_r$ is 3-colourable by Lemma \ref{unq3col},
then the last deleted point is added to it and a constant number of possible 4-colourings are constructed 
and checked due to Lemma \ref{cont4col}.
Each 4-colouring is considered one by one, and each of the deleted point is added in the reversed order of its deletion,
and assigned the unique colour determined by the $K_3$ it sees in the remaining point set. At each step, it is checked whether the 4-colouring is valid or not. If at any step
two mutually visible points are forced to have the same colour, then $P$ does not have a 4-colouring, with the chosen 4-colouring of $P_r$. Otherwise, after adding all the 
deleted points, a 4-colouring is obtained in $O(n^3)$ time.
$\\ \\$
If the reduced set $P_r$ is not 3-colourable, then three distinct colours are assigned to its convex hull vertices, and all eight possible 4-colourings are found in $O(n^2)$ time due to Lemma \ref{fixcol}.
It is checked whether any of these colourings is a valid 4-colouring or not. If some valid 4-colourings are obtained, then each of them is considered one by one,
and as before, the deleted points are added and coloured.  If at any step
two mutually visible points are forced to have the same colour, then $P$ does not have a 4-colouring, with the chosen 4-colouring of $P_r$. Otherwise, after adding all the deleted points, 
a 4-colouring is obtained in $O(n^3)$ time.
The whole algorithm takes $O(n^4)$ time.
$\\ \\$
\qed
\section{$5$-colouring point visibility graphs} \label{sec5col}

In this section we prove that deciding whether a PVG with a given embedding is $5$-colourable, is NP-hard.
We provide a reduction of 3-SAT to the PVG $5$-colouring problem.
We use the reduction of 3-SAT to the $3$-colouring problem of general graphs.
\subsection{$3$-colouring a general graph} 
\begin{figure}
\begin{center}
\centerline{\hbox{\psfig{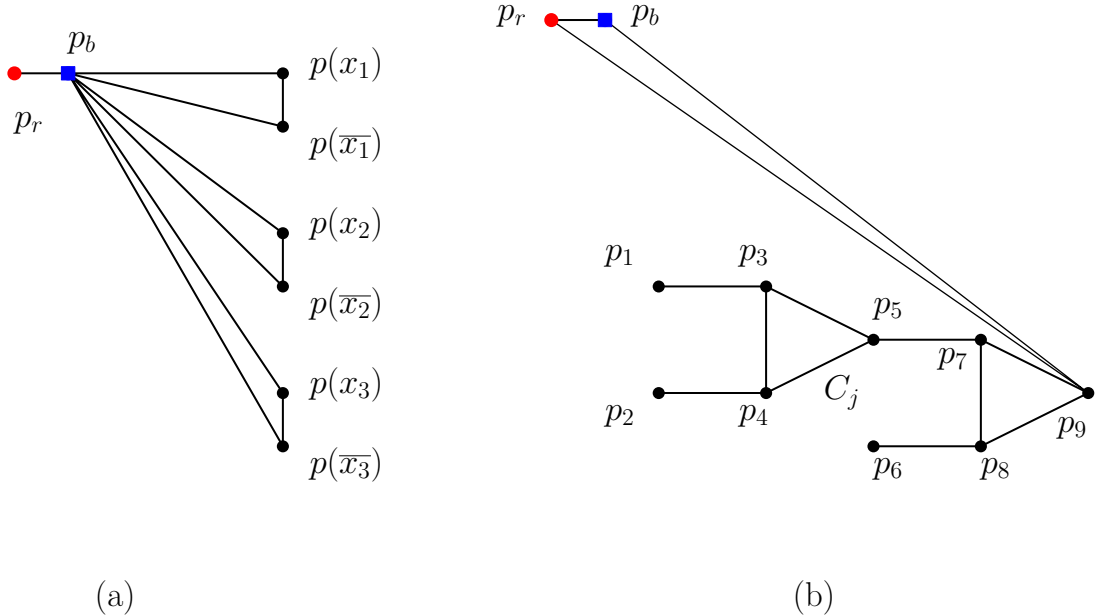}}}
\caption{(a) A variable gadget. (b) A clause gadget.}
\label{gadgets}
\end{center}
\end{figure}
\begin{figure}
\begin{center}
\centerline{\hbox{\psfig{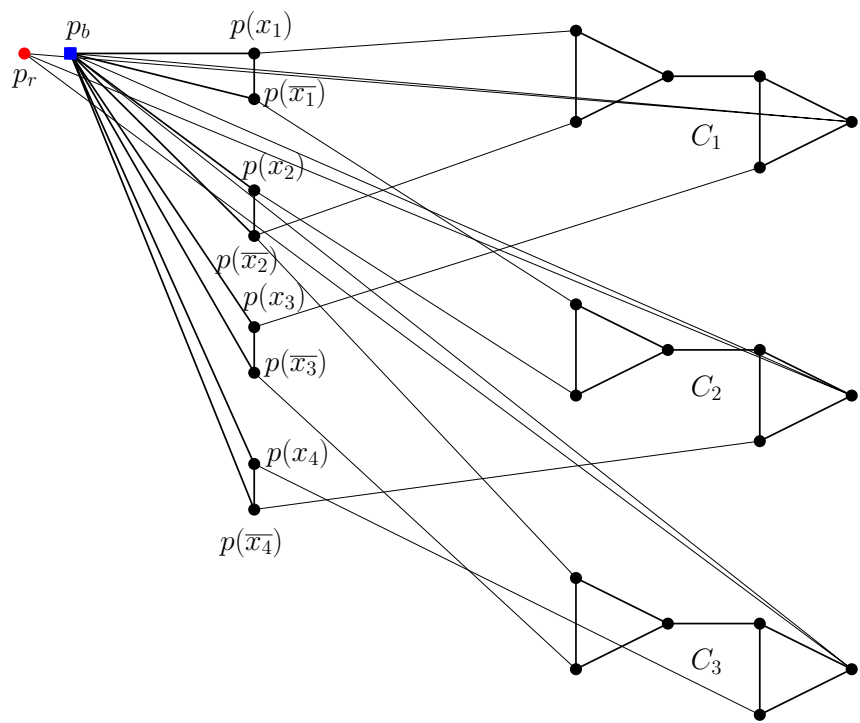}}}
\caption{The full embedding for a given 3-SAT formula.}
\label{gengraph}
\end{center}
\end{figure}

For convenience, we first briefly describe the reduction for the $3$-colouring of general graphs, considering
the graph as an embedding $\xi$ of points and line segments in the plane \cite{cai-79}.
Consider a 3-SAT formula $\theta$ with variables $x_1, x_2, \ldots, x_n$ and clauses $C_1, C_2, \ldots, C_m$.
Suppose the corresponding graph is to be coloured with red, green and blue.
Consider Figure \ref{gadgets}(a). It shows the variable-gadgets. The points representing a variable $x_i$ and its negation (say, $p(x_i)$ 
and $p(\overline{x_i})$, respectively)
are adjacent to each other, making $n$ pairs altogether. No two points in different pairs are adjacent to each other.
A separate point $p_b$ is wlog assumed to be blue and made adjacent to all the other points in the variable gadgets. So, each variable
point must have exactly one red and one green point. For variable points, let green and red represent an assignment of 
true and false, respectively. The point $p_b$ is also adjacent to a separate point $p_r$ assumed to be red.
$\\ \\$
Now consider Figure \ref{gadgets}(b). It shows a clause gadget. Suppose that the points $p_1$, $p_2$ and $p_6$ can be coloured
only with green and red. Then $p_9$ can be coloured with green if and only if at least one of $p_1$, $p_2$ and $p_6$
are coloured with green. To prevent $p_9$ from being coloured red or blue, $p_9$ is made adjacent to $p_r$ and $p_b$.
$\\ \\$
The whole embedding corresponding to the 3-SAT formula is shown in Figure \ref{gengraph}. The points $p_r$ and $p_b$ are the 
same for all variables and clauses. 
For each clause gadget, the points corresponding to  $p_1$, $p_2$ and $p_6$ are in the respective variable gadgets.
Thus, for $n$ variables and $m$ clauses, $\xi$ has $2n + 6m + 2$ points in total.
\subsection{Transformation to a point visibility graph}
\begin{figure}
\begin{center}
\centerline{\hbox{\psfig{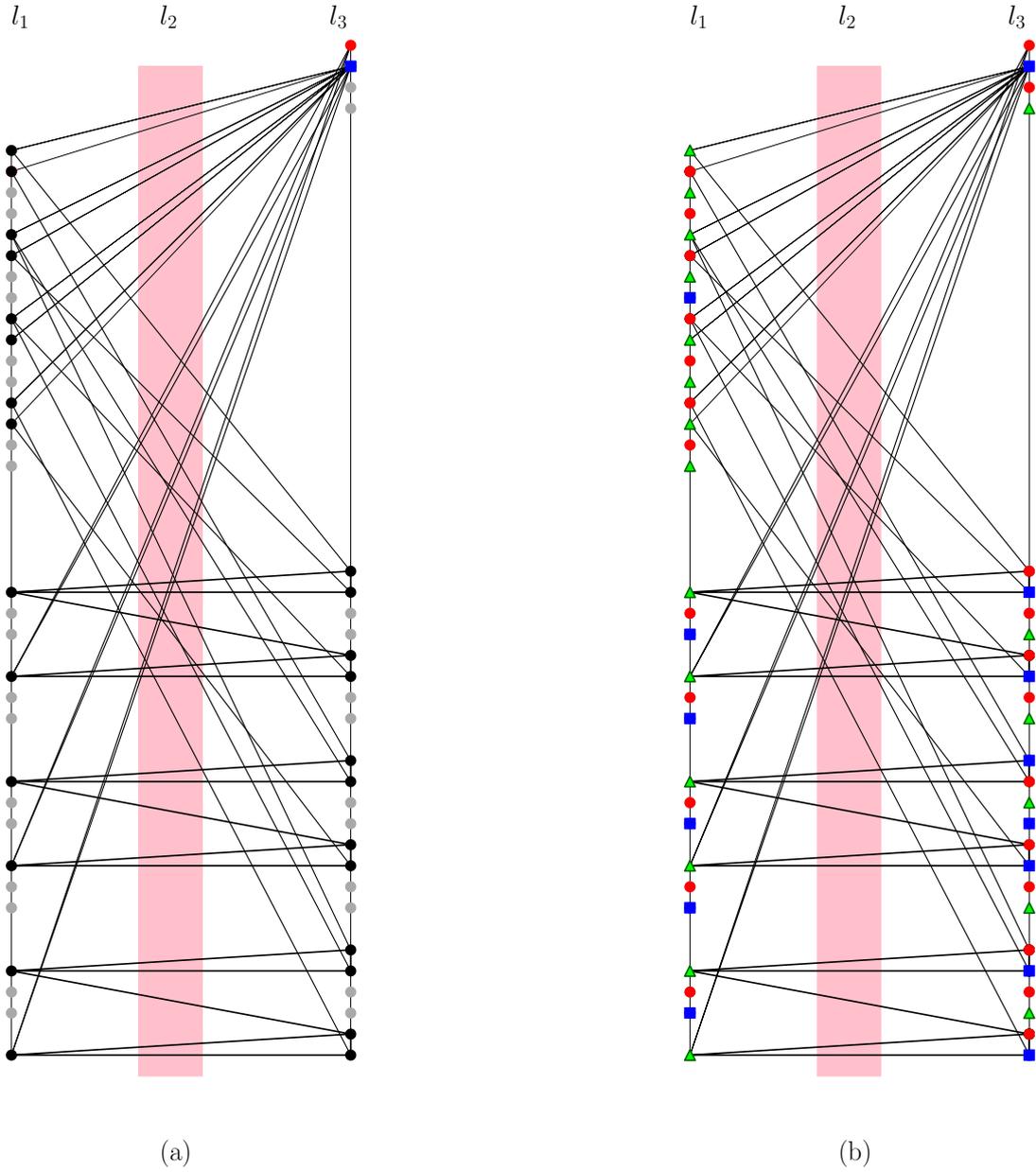}}}
\caption{(a) The new embedding $\zeta$ on three vertical lines. The dummy points are shown in gray. (b) A 3-colouring of $\zeta$ representing $x_1, x_2, \overline{x_3}$
and $\overline{x_4}$ assigned $1$ in $\theta$.}
\label{finalemb}
\end{center}
\end{figure}

Consider the following transformation of $\xi$ into a new embedding $\zeta$ (Figure \ref{finalemb}(a)) for a given 3-SAT formula $\theta$.
We use some extra points called \emph{dummy points} to act as blockers during the transformation.
\begin{enumerate} [(a)]
 \item All points of $\xi$ are embedded on two vertical lines $l_1$ and $l_3$.
 \item Two points $p_r$ and $p_b$ are placed on $l_3$ above all other points of $\xi$, followed by a dummy point.
 \item Each pair of variable gadget points are embedded as consecutive points on $l_1$. 
 \item Separating a variable gadget from the next variable gadget is a dummy point.
 \item For every clause gadget, the points corresponding to $p_5$ and $p_9$ are on $l_1$,
 separated by a dummy point.
 \item For every clause gadget, the points corresponding to $p_3$, $p_4$, $p_7$ and $p_8$ 
 are on $l_3$, in the vertical order from top to bottom. The points $p_3$ and $p_4$ are consecutive.
 The points $p_7$ and $p_8$ are consecutive. There is a dummy point between $p_4$ and $p_7$.
 \item The points of consecutive clause gadgets are separated by a dummy point each.
 \item Let $l_2$ be a vertical line lying between $l_1$ and $l_3$. On $l_2$, embed points to block all visibility relationships other than those corresponding to edges in $\xi$.
 Perturb the points of $l_1$ and $l_3$ so that each point in $l_2$ blocks exactly one pair of points.
\end{enumerate}
The total number of points needed in the new embedding is as follows:
\begin{itemize}
 \item $p_r$ and $p_b$ are $2$ points.
 \item Variable gadgets are $2n$ points.
 \item Clause gadgets are $2m$ points on $l_1$ and $4m$ points on $l_3$.
 \item There are $n + 2m - 1$ dummy points on $l_1$ and $2m$ dummy points on $l_3$.
 \item Thus, there are $3n + 4m - 1$ points on $l_1$ and $6m +2$ points on $l_3$. 
 \item There are $9m + 2n$ edges from $\xi$ between $l_1$ and $l_3$.  
 \item Thus, there are $(3n + 4m - 1)(6m +2) - (9m + 2n)$ points on $l_2$ to block the visibility of the rest of the pairs.
\end{itemize}
 \lemma \label{poly}
 The above construction can be achieved in polynomial time.
 
\proof
 As shown above, the number of points used is polynomial. 
All the points of $l_1$ and $l_3$ are embedded on lattice points. 
The intersections of the line segments joining points of $l_1$ and $l_3$
are computed, and $l_2$ is chosen such that none of these intersection point lies on $l_2$.
 To block visibilities, 
 and the intersection of the line segments joining points of $l_1$ and $l_3$ with $l_2$ are computed and blockers are placed on the intersection
 points. All of this is achievable in polynomial time.
\qed
 \lemma \label{iff}
 The PVG of $\zeta$ can be $5$-coloured if and only if $\theta$ has a satisfying assignment.
 
\proof Suppose that $\theta$ has a satisfying assignment. Then the points of $\zeta$ obtained from $\xi$ can be coloured with red, blue 
 and green. 
 The two neighbours of a dummy point on $l_1$ or $l_3$ can have at most two colours, so the dummy point can always be assigned the third colour. 
 The points on $l_2$ are coloured alternately with two different colours
 (Figure \ref{finalemb}(b)).
 $\\ \\$
 Now suppose that $\zeta$ has a 5-colouring. All points of $l_2$ are visible from all points of $l_1$ and $l_3$. The points of 
 $l_2$ must be coloured at least with two colours. This means, the points of the graph induced by $\xi$ are coloured with 
 at most $3$ colours, which is possible only when $\theta$ is satisfiable.
\qed
We have the following theorem.
\theorem
The problem of deciding whether the 
 visibility graph of a given point set is $5$-colourable, is NP-complete.

\proof
 A $5$-colouring of a point visibility graph can be verified in polynomial time. Thus, the problem is in NP.
 On the other hand, 3-SAT can be reduced to the problem.
 Given a 3-SAT formula $\theta$, by Lemmas \ref{poly} and \ref{iff}, a point set $\zeta$ can be constructed in time polynomial in the size of $\theta$
 such that $\zeta$ can be $5$-coloured if and only if $\theta$ has a satisfying assignment. Thus, the problem is NP-complete.
\qed

\section{Colouring a point set with small clique number}
In general, graphs with small clique numbers can have arbitrarily large chromatic numbers. 
In fact, there exist triangle free graphs with arbitrarily high chromatic numbers due to the construction of Mycielski \cite{myc}. 
Pfender showed that for a PVG with $\omega (G) = 6$, $\chi (G)$ can be arbitrarily large \cite{p-vgps-2008}.
But it is not known whether the chromatic number of a PVG is bounded, if its clique number is only $4$ or $5$.
Here, we address this question.
\subsection{A graph with $\omega (G) = 4$ and $\chi (G) = 6$} \label{secexample2}
\begin{figure}
\begin{center}
\centerline{\hbox{\psfig{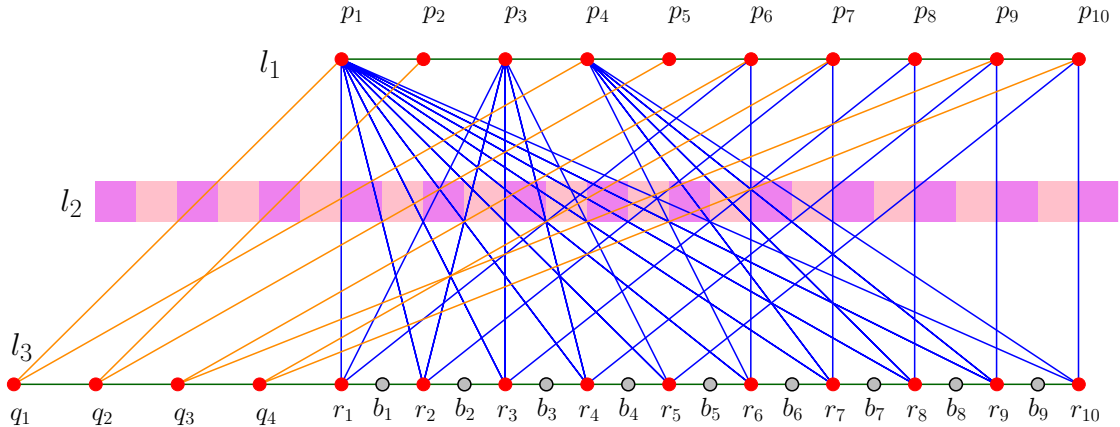}}}
\caption{A point visibility graph with clique number four but chromatic number six.}
\label{6graph}
\end{center}
\end{figure}

 K\'{a}ra et al \cite{kpw-ocnv-2005} showed that a PVG with clique number $4$ can have chromatic number $5$.
They then generalized their example to prove that there is an exponential function $f$ such that for a family
of PVGs, the identity $\chi (G) = f(\omega (G))$ holds for all graphs in the family.
The main question remaining is whether PVGs with maximum clique size $4$ have bounded chromatic
number.  
Here we construct a visibility graph $G'$ with $\omega (G') = 4$ and $\chi (G') = 6$ (Figure \ref{6graph}).
We construct $G'$ directly as a visibility embedding, as follows.
\begin{enumerate}
 \item  Consider three horizontal lines $l_1$, $l_2$ and $l_3$ parallel to each other.
\item On the first line, embed ten points $\{ p_1, p_2, \ldots, p_{10}\}$ from left to right. 
\item From left to right embed the points $q_1, \ldots, q_4$ on $l_3$.
\item Join with line segments the pairs $(q_1,p_1)$, $(q_1,p_4)$, $(q_2,p_2)$, $(q_2,p_5)$. \label{step1}
\item Join with line segments the pairs $(q_3,p_6)$, $(q_3,p_9)$, $(q_4p_7)$, $(q_4,p_{10})$.
\item Starting from the right of $q_4$, embed the points $r_1, b_1, r_2, b_2, \ldots, r_{10}$ on $l_3$.
\item Join each $r_i$ with $1 \leq i \leq 5$ with $p_1$, $p_3$ and $p_{i+5}$. 
\item Join each $r_i$ with $6 \leq i \leq 10$ with $p_1$, $p_4$ and $p_{i}$. \label{step2}
\item Embed points on $l_2$ such that only the adjacencies described from steps \ref{step1} to \ref{step2} hold.
\end{enumerate}
We have the following lemmas.
 \lemma
 The clique number of $G'$ is $4$.
 
\proof
By construction, the points on $l_1$ and $l_3$ together induce a triangle free graph. The points of $l_2$ can contribute
at most two more points to cliques induced by the points on $l_1$ and $l_3$. So, $G'$ has cliques of size at most four.
\qed

 \lemma
 The chromatic number of $G'$ is $6$.
 
\proof
 Each point on $l_2$ is adjacent to every point on $l_1$ and $l_3$, so the points on $l_2$ require two colours which are absent from the points
 of $l_1$ and $l_3$. So, it suffices to show that the graph induced by points on $l_1$ and $l_3$ is not three colourable. Suppose that the points
 $p_1, \ldots p_5$ have only two colours (say, $C_1$ and $C_2$). This means that they are coloured with $C_1$ and $C_2$ alternately, 
 and $q_1$ and $q_2$ must be coloured with two extra colours, $C_3$ and $C_4$ respectively.
 $\\ \\$
 Now suppose that all three of $C_1$, $C_2$ and $C_3$ occur among $p_1, \ldots p_5$. Similarly, 
 all three of $C_1$, $C_2$ and $C_3$ occur among $p_6, \ldots p_{10}$, for otherwise the previous argument is applicable to $q_3$, $q_4$
 and $p_6, \ldots p_{10}$. Also, $p_1$ and $p_3$, or $p_1$ and $p_4$ must have two distinct colours. On the other hand, three distinct colours 
 must also occur among $p_6, \ldots p_{10}$. But for every $p_i$, $6 \leq i \leq 10$, there are two points among $r_1, r_2, \ldots, r_{10}$
 that are adjacent to $p_i$ and one pair among $\{p_1, p_3\}$ and $\{p_1, p_4\}$. So, at least one of the points among  
 $r_1, r_2, \ldots, r_{10}$ must have the fourth colour.
\qed

\section{Concluding Remarks} 
We have settled the question of colouring PVGs, showing that the $4$-colour problem on them is solvable in polynomial time while the $5$-colour
problem is NP-complete.
We have shown that there is a PVG with clique number four but chromatic number six. However, it is still open to show whether a PVG with clique number
four can have a greater chromatic number or not. 
\bibliographystyle{plain}
\bibliography{vis}

\end{document}